\documentclass[proof]{pasj01}
\usepackage{graphicx}

\Received{}
\Accepted{}

\SetRunningHead{Kaneko~et~al.}{Physical states of molecular gas in the overlapping region of interacting galaxies}

\begin{document}
	\title{Investigating physical states of molecular gas in the overlapping region of interacting galaxies NGC\,4567/4568 using ALMA}	\author{
		Hiroyuki \textsc{Kaneko},\altaffilmark{1,2,*}
		Shoya \textsc{Tokita},\altaffilmark{3}
		and Nario \textsc{Kuno}\altaffilmark{3,4}
	}
	\altaffiltext{1}{Graduate School of Education, Joetsu University of Education, 1 Yamayashiki-machi, Joetsu, Niigata, 943-8512, Japan}
	\altaffiltext{2}{National Astronomical Observatory of Japan, 2-21-1 Osawa, Mitaka, Tokyo, 181-8588, Japan}
	\altaffiltext{3}{Graduate School of Pure and Applied Sciences, University of Tsukuba, 1-1-1 Tennodai, Tsukuba, Ibaraki, 305-8577, Japan}
	\altaffiltext{4}{Tomonaga Center for the History of the Universe, University of Tsukuba, Tsukuba, Ibaraki 305-8571, Japan}
	\email{kaneko.hiroyuki.astro@gmail.com}
	
	
	\maketitle

\begin{abstract}
	We present ALMA observations of a diffuse gas tracer, CO($J$ = 1--0), and a warmer and denser gas tracer, CO($J$ = 3--2), in the overlapping region of interacting galaxies NGC\,4567/4568, which are in the early stage of interaction. 
	To comprehend the impact of galaxy interactions on molecular gas properties, we focus on interacting galaxies during the early stage and study their molecular gas properties. 
	In this study, we investigate the physical states of a filamentary molecular structure at the overlapping region, which was previously reported.
	Utilising new higher-resolution CO($J$ = 1--0) data, we identify molecular clouds within overlapping and disc regions.
	Although the molecular clouds in the filament have a factor of two higher an average virial parameter (0.56\,$\pm$\,0.14) than that in the overlapping region (0.28\,$\pm$\,0.12) and in the disc region (0.26\,$\pm$\,0.16), all identified molecular clouds are gravitationally bound.
	These clouds in the filament also have a larger velocity dispersion than that in the overlapping region, suggesting that molecular gas and/or atomic gas with different velocities collide there.
	We calculate the ratio of the integrated intensity of CO($J$ = 3--2) and CO($J$ = 1--0) (= $R_{3-2/1-0}$) on the molecular cloud scale.
	The maximum $R_{3-2/1-0}$ is 0.17\,$\pm$\,0.04 for all identified clouds.
	The $R_{3-2/1-0}$ of the molecular clouds in the filament is lower than that of the surrounding area.
	This result contradicts the predictions of previous numerical simulations, which suggested that the molecular gas on the collision front of galaxies is compressed and becomes denser. 
	Our results imply that NGC\,4567/4568 is in a very early stage of interaction; otherwise, the molecular clouds in the filament would not yet fulfil the conditions necessary to trigger star formation.
\end{abstract}


\section{Introduction}
Galaxy interactions (fly-bys, collisions, and mergers) play key roles in drastically altering the distribution and kinematics of stars and gas.
As such, investigating the details of interacting galaxies is essential for comprehending the evolution and diversity of galaxies.
One of the most important characteristics of interacting galaxies is an enhancement of star formation activity (e.g., \cite{Bushouse86}).
The morphology of interacting galaxies during the early stages of interaction is relatively undisturbed, and the star formation rate (SFR) is only a few times higher than that of isolated galaxies (e.g., \cite{Knapen09}).
On the other hand, in the late stages of interaction, the merger of two galactic nuclei can lead to SFRs that are more than 10 times higher \citep{Kennicutt87,Teyssier10}. 
Despite this, the mechanism behind active star formation in interacting galaxies remains uncertain.
In particular, interpreting the mechanism of star formation in late-stage mergers (e.g., ultra-luminous infrared galaxies: \cite{Sanders96}) is significantly hampered by the complex and disturbed morphology resulting from tails and arms emanating from the host galaxies. 
In contrast, interacting galaxies in the early stages of interaction are relatively undisturbed, preserving the original morphology and properties of the interstellar medium (ISM) to a reasonable degree.
This characteristic of interacting galaxies in the early stages of interaction allows for detailed studies of the spatial distribution of the conditions of the ISM and its relation to the star formation activity triggered by the tidal interaction between two disks.

Since molecular gas is fuel for star formation, investigating how the interaction affects molecular gas properties is important for understanding the mechanism of intense star formation in interacting galaxies.
Observations have revealed that a large amount of molecular gas is accumulated in the overlapping region of interacting galaxies \citep{Vollmer21,Wilson03,Zhu07}. 
In particular, the Antennae Galaxies, known as interacting galaxies during the mid-stages of interaction due to their severely disturbed morphologies, possess filamentary giant molecular clouds (with lengths of a few kiloparsecs) in the overlapping region between the two galaxies \citep{Wilson00,Whitmore14}.
In the overlapping region, many starbursts and super star clusters (SSCs) are found, leading to the supposition that SSCs are formed from such large and massive giant molecular clouds \citep{Wilson03,Mengel08}. 
Recent numerical simulations demonstrate that the shock induced by the collision compresses molecular gas, converting it into denser gas, thereby creating giant molecular clouds with a filamentary structure similar to that seen in the Antennae Galaxies at the overlapping region during the first encounter \citep{Saitoh09}.
Soon after the formation of the filament, it fragments into smaller molecular clouds, and many SSCs are formed from these clouds, resulting in starbursts \citep{Saitoh11}.
These studies suggest that SSCs are produced through collisions.
As numerical simulations have demonstrated that the filament can form in the early stages of interaction, significant changes in the properties of molecular gas, the fuel for stars, are expected to occur even during the early stage.
Therefore, it is important to study the molecular gas of interacting galaxies during the early stages of the interaction. 
Since only a limited number of observational studies have been focused on this aspect, our study can provide new insights into the mechanisms of tidally-driven star formation.

The interacting galaxy pair in the Virgo Cluster NGC\,4567/4568 is one of the best targets for investigating tidally-driven star formation. 
The galaxy pair is likely in the early stage of collision and probably experiencing its first encounter \citep{Kaneko13}, as evidenced by its weakly disturbed morphology and relatively large projected nuclear separation (5.6\,kpc). 
Furthermore, the close proximity (16 Mpc: \cite{Mei07}) and small inclination of both constituent galaxies (\timeform{44D} for NGC\,4567 and \timeform{58D} for NGC\,4568) enable us to study the details of the inner structure at a high linear resolution.
The basic properties of the galaxy pair are summarised in table \ref{Table1}.
The molecular gas fraction (the fraction of molecular hydrogen in the sum of the surface density of molecular and atomic hydrogen) in the overlapping region is 5-15\% higher than that in the southern disc of NGC\,4568, which has the same surface density of total (i.e., H$_{2}$+H\emissiontype{I}) gas.
Since the surface density of total gas is comparable to the opposite side of the disc with the same radius, the higher molecular gas fraction in the overlapping region may not be explained by selective H\emissiontype{I} stripping by the cluster.
This fact indicates that the conversion from H$\emissiontype{I}$ to H$_{2}$ may have already occurred \citep{Kaneko17}. 
These observational findings suggest that active star formation induced by galaxy interaction can be expected since the amount of molecular gas is increased.
They also suggest giant molecular clouds may be efficiently formed in the overlapping region of this galaxy pair.
Furthermore, \citet{Kaneko18} discovered a long filamentary molecular structure, which is thought to be a molecular collision front. 
The filament is similar to the structure predicted by numerical simulations \citep{Saitoh09}.
By investigating the properties of the filament formed by the galaxy collision, we intend to gain insight into understanding how off-centre starbursts (e.g., Antennae Galaxies) are caused in interacting galaxies. 

\begin{table}
	\centering
	\tbl{Basic parameters of NGC\,4567 and NGC\,4568.}{
		\begin{tabular}{lcc} 
			\hline
			& NGC\,4567 & NGC\,4568\\
			\hline
			Morphological type\footnotemark[*] & SA(rs)bc & SA(rs)bc \\
			Right Ascension (J2000.0)\footnotemark[$*$] & \timeform{12h36m32.7s} & \timeform{12h36m34.2s} \\
			Declination (J2000.0)\footnotemark[$*$] & \timeform{11D15'29.0"} & \timeform{11D14'20.0"} \\
			Distance\footnotemark[$\dag$] & \multicolumn{2}{c}{16 Mpc} \\
			${V_{\mathrm{LSR}}}$\footnotemark[$*$] & 2247 km s$^{-1}$ & 2255 km s$^{-1}$ \\
			Inclination angle\footnotemark[$*$] & \timeform{44D} & \timeform{58D} \\
			Linear scale & \multicolumn{2}{c}{77.5 pc arcsec$^{-1}$} \\
			\hline
		\end{tabular}}
	\label{Table1}
	\begin{tabnote}
		\footnotemark[*] NASA/IPAC Extragalactic Database (NED)\\
		\footnotemark[$\dag$] \citet{Mei07}
	\end{tabnote}
\end{table}

Observations have shown that HCN($J$ = 1--0)/CO($J$ = 1--0) ratio is enhanced in galaxies with high star formation efficiency, like luminous infrared galaxies \citep{Gao04,Gracia08,Garcia12}. 
However, HCN($J$ = 1--0) line is usually fainter than CO($J$ = 1--0) by more than an order of magnitude. 
Therefore, strong HCN emission is not expected for this galaxy pair, as the starburst has not been triggered yet.
On the other hand, low-$J$ CO molecules are known to be easily excited and suitable tracers for understanding the physical states of molecular gas.
Due to its higher excitation energy ($T_{\rm ex}$ = 33 K) and critical density ($>10^{4}$ cm$^{-3}$), CO($J$ = 3--2) traces warmer and denser molecular gas than CO($J$ = 1--0).
This means that the CO($J$ = 3--2) line is more directly related to star-forming regions than the CO($J$ = 1--0) line.
In addition, the critical density of CO($J$ = 3--2) is an order of magnitude lower than that of HCN.
Therefore, we can investigate the dense gas formation process by comparing the spatial distribution of CO($J$ = 3--2) with that of CO($J$ = 1--0).
For these reasons, we conducted CO($J$ = 1--0) and CO($J$ = 3--2) line observations of NGC\,4567/4568 using ALMA and discussed the molecular gas properties in the overlapping region of this interacting pair. 
We identified molecular clouds in both CO($J$ = 1--0) and CO($J$ = 3--2), and derived $R_{3-2/1-0}$ = $I_{\mathrm{CO}(J=3-2)}/I_{\mathrm{CO}(J=1-0)}$ ($I_{\mathrm{CO}(J=3-2)}$ and $I_{\mathrm{CO}(J=1-0)}$ are intensities of CO($J$ = 3--2) and CO($J$ = 1--0), respectively) of each molecular cloud to investigate whether the clouds are dense enough to trigger star formation.

This paper is organised as follows:
Section 2 describes the ALMA observations and data reduction. 
In section 3, we show the results of distributions of CO($J$ = 1--0) and CO($J$ = 3--2). 
In section 4, we identify the molecular clouds and investigate their physical properties in each region, including the filament, which is considered a molecular collision front. 
Finally, we summarise our results in section 5.

\section{Observations and data reduction}
\begin{figure}
	\begin{center}
		\includegraphics[width=\linewidth]{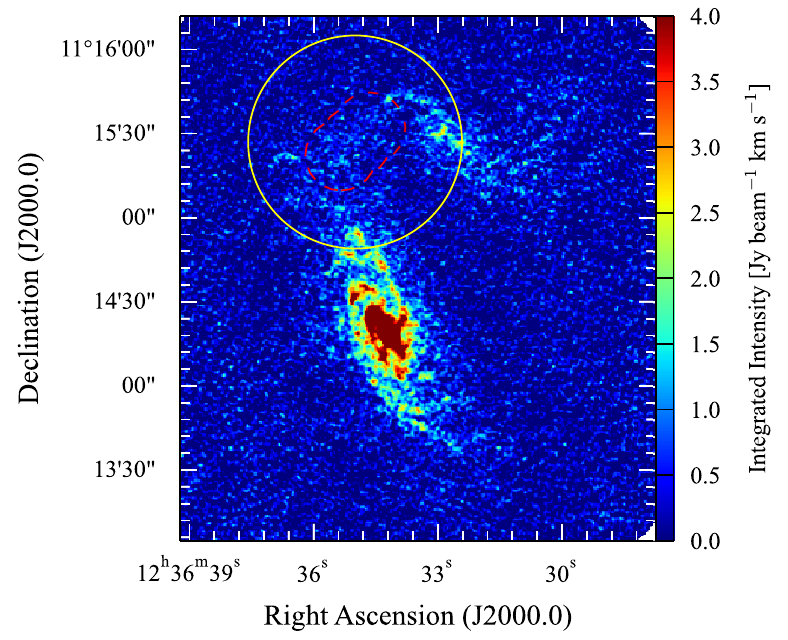}
	\end{center}
	\caption{Field of View of ALMA Cycle 3 observations superposed on CO($J$ = 1--0) integrated intensity map obtained with the ALMA Cycle 1 \citep{Kaneko18}. NGC\,4567 is on the top right, while NGC\,4568 is on the bottom left. Yellow-solid circle: Field of view of CO($J$ = 1--0). Red-dashed region: Field of view of CO($J$ = 3--2).}
	\label{cycle3_fov}
\end{figure}

We observed NGC\,4567/4568 using ALMA as a Cycle 3 program (ID: 2015.1.01161.S, PI: Hiroyuki Kaneko).
Forty-nine 12-m array and nine Atacama Compact Array (ACA) 7-m array antennas were used for CO($J$ = 1--0) observations with the Band 3 receiver, while thirty-seven 12-m antennas and eight ACA 7-m antennas were used for CO($J$ = 3--2) observations with the Band 7 receiver.
The primary beam sizes for the 12-m array and 7-m array at 115.271 GHz (the rest frequency of CO($J$ = 1--0)) are 50.5 and 86.6 arcsec, respectively, and those at 345.796 GHz (the rest frequency of CO($J$ = 3--2)) is 16.8 and 28.9 arcsec.
The CO($J$ = 1--0) data were obtained in a single field of view, whereas the CO($J$ = 3--2) data were obtained in a mosaic of three fields of view (figure \ref{cycle3_fov}).
The CO($J$ = 1--0) observations were conducted in January, May and June 2016 with a total observation time of 3.71 hours, and the CO($J$ = 3--2) observations were carried out in May--July 2016 with a total observation time of 0.52 hours. 
The bandpass and phase for CO($J$ = 1--0) and CO($J$ = 3--2) observations were calibrated using J1229+0203 and J1239+0730, respectively.

Data reduction, including calibration and imaging, was executed with the Common Astronomy Software Applications (CASA) package \citep{CASA22}.
The 12-m array data and the ACA 7-m array data for CO($J$ = 1--0) were combined together.
Initially, the 3-mm and 1-mm continuum emission were scrutinized and found to be undetected.
The 3\,$\sigma$ upper limits for the 3-mm and 1-mm continuum were 1.7\,$\times$\,10$^{-1}$ mJy beam$^{-1}$ (with a beam size of \timeform{2.0''}\,$\times$\,\timeform{1.6''}) and 2.4\,$\times$\,10$^{-1}$ mJy beam$^{-1}$ (with a beam size of \timeform{0.66''}\,$\times$\,\timeform{0.53''}), respectively.
To image the CO lines, we performed continuum subtraction using the {\itshape uvcontsub} task and subsequently corrected for primary beam attenuation.
The imaging process was conducted interactively using the {\itshape tclean} task with Briggs weighting (using a robustness parameter of 0.5).
The systematic error on the absolute flux is estimated to be approximately 5 and 10\% for both sidebands in Band 3 and Band 7, respectively.

The final data of CO($J$ = 1--0) were imaged on a 256 $\times$ 256 pixel with a grid size of \timeform{0.3"}.
The combined cube has an angular resolution of \timeform{1.62"}\,$\times$\,\timeform{1.28"} (126 pc$\times$99 pc at 16 Mpc) with a position angle of \timeform{115D} and a velocity resolution of 5 km s$^{-1}$.
As a result, the root mean square (rms) noise of 1.5 mJy beam$^{-1}$ (0.067 mK) was obtained.
Thus, our CO($J$ = 1--0) data has a higher angular resolution by a factor of 1.5 and higher sensitivity by a factor of three when compared to the Cycle 1 data (an angular resolution of \timeform{2.0''}\,$\times$\,\timeform{2.0''} and the rms noise of 8.5 mJy beam$^{-1}$, corresponding to 0.20 mK: \cite{Kaneko18}).
Additionally, we estimated the missing flux for CO($J$ = 1--0) in Cycle 3 using data obtained with the Nobeyama 45-m telescope \citep{Kaneko13}, and found that the missing flux accounts for 63\% of the total flux.

CO($J$ = 3--2) were imaged on a 500\,$\times$\,500 pixel with a grid size of \timeform{0.08"}.
The resulting combined data cube has an angular resolution of \timeform{0.56"}\,$\times$\,\timeform{0.47"} (43 pc\,$\times$\,36 pc at 16 Mpc), a position angle of \timeform{119D}, and a velocity resolution of 5 km s$^{-1}$.
The rms noise for CO($J$ = 3--2) is 1.5 mJy beam$^{-1}$. 
As no single-dish data was available, the missing flux of CO($J$ = 3--2) could not be estimated.
	
\section{Results}
\subsection{CO($J$ = 1--0)}\label{CO10}
\begin{figure*}
	\begin{center}
		\includegraphics[width=\linewidth]{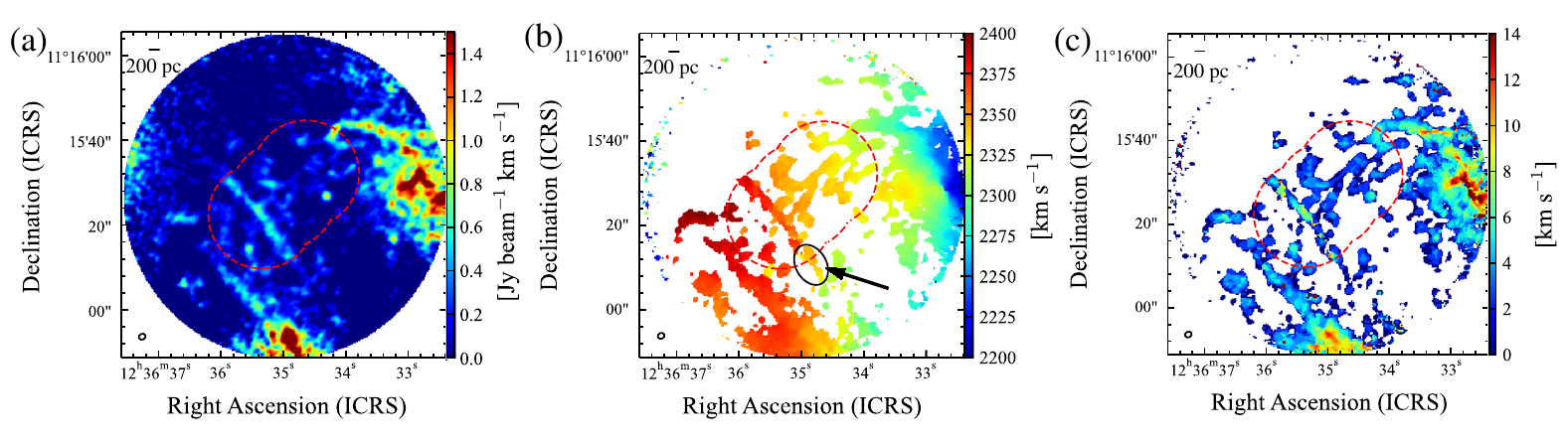}
	\end{center}
	\caption{CO($J$ = 1--0) moment maps. (a) Integrated Intensity map. (b) Velocity field. (c) Velocity dispersion map. The beam size of \timeform{1.62"}\,$\times$\,\timeform{1.28"} and the scale bar of 200 pc are shown in the bottom-left corner and top-left corner of each map. Red-dashed regions indicate the field of view of CO($J$ = 3--2). Molecular gas indicated with black eclipse and arrow in (b) may not consist the filament (see, section \ref{CO10} and \ref{sec:pvd}).}
	\label{co10_moment_map}
\end{figure*}

The moment maps of CO($J$ = 1--0) are shown in figure \ref{co10_moment_map}.
The velocity field and velocity dispersion maps were clipped at 5\,$\sigma$ (1\,$\sigma$ = 1.5 mJy beam$^{-1}$).
The velocity field and velocity dispersion maps, which were not presented in the previous work by \citet{Kaneko18}, provide a more detailed depiction of the gas kinematics in the overlapping region at sub-kiloparsec resolution.
	
In figure \ref{co10_moment_map}(a), the molecular discs of NGC\,4567 and NGC\,4568 are seen on the western and southern sides, respectively. 
At the centre of the field of view, the filamentary molecular structure, which was discovered by \citet{Kaneko18}, is visible.
We have identified some clumpy molecular clouds in the field of view of Band 7 observations (the red-dashed region in figure \ref{co10_moment_map}). 
Since the velocity field (figure \ref{co10_moment_map}(b)) is not significantly disturbed in the overlapping region, it suggests that the two galaxies have undergone their initial encounter very recently.
The primary velocity component of the southwest part of the ``filament'' exhibits a value of $\sim$2325 km s$^{-1}$ (indicated with a black eclipse and an arrow in figure \ref{co10_moment_map}(b)), whereas the rest of the filament has a velocity component of $\sim$2375 km s$^{-1}$.  
The two possibilities can explain this difference:
(1) the southwest component is not a part of the filament, or (2) the component is not part of the filament but rather an extension of the spiral arm from NGC\,4567.
This issue will be discussed in section \ref{sec:pvd}.
A smooth gradient of the velocity field illustrates why previous low angular- and velocity-resolution observations before \citet{Kaneko18} were unable to reveal the filament.
Figure \ref{co10_moment_map}(c) shows that the velocity dispersion is higher in the filament, suggesting that molecular gas collision occurs there, as indicated in \citet{Kaneko18}.
	
The CO($J$ = 1--0) integrated intensity in the field of view is 205\,$\pm$\,3 Jy km s$^{-1}$.
We derived the luminosity mass of the observed region from the CO($J$ = 1--0) flux using the following equation \citep{Kenney98}:
\begin{equation}
	\left(\frac{M_\mathrm{H_2}}{M_\odot}\right) = 3.9\times10^{-17}\left(\frac{X_\mathrm{CO}}{\mathrm{cm}^{-2} (\mathrm{K km s}^{-1})^{-1}}\right)\left(\frac{D}{\mathrm{Mpc}}\right)^{2}\left(\frac{S_\mathrm{CO}}{\mathrm{Jy km s^{-1}}}\right), \label{H2mass}
\end{equation}
where $M_\mathrm{H_2}$ is the molecular gas mass, $X_\mathrm{CO}$ is a CO-H$_2$ conversion factor, and $D$ is the distance to the galaxy.
Adopting $X_{\rm CO}$ = 2.0\,$\times$\,10$^{20}$ [cm$^{-2}$ (K km s$^{-1})^{-1}$] \citep{Bolatto13}, the luminosity mass is (4.1\,$\pm$\,0.1)\,$\times$\,10$^{8}$ $M_\odot$.
Since the total amount of molecular gas of the NGC 4567/4568 pair is (4.7\,$\pm$\,0.7)\,$\times$\,10$^{9}$ $M_\odot$ \citep{Kaneko13}, taking into account the difference of $X_{\rm CO}$, this corresponds to 9\% of molecular gas of this pair.
Note that the derived mass is the lower limit due to missing flux.
	
\subsection{CO($J$ = 3--2)}
\begin{figure*}[bt!]
	\begin{center}
		\includegraphics[width=\linewidth]{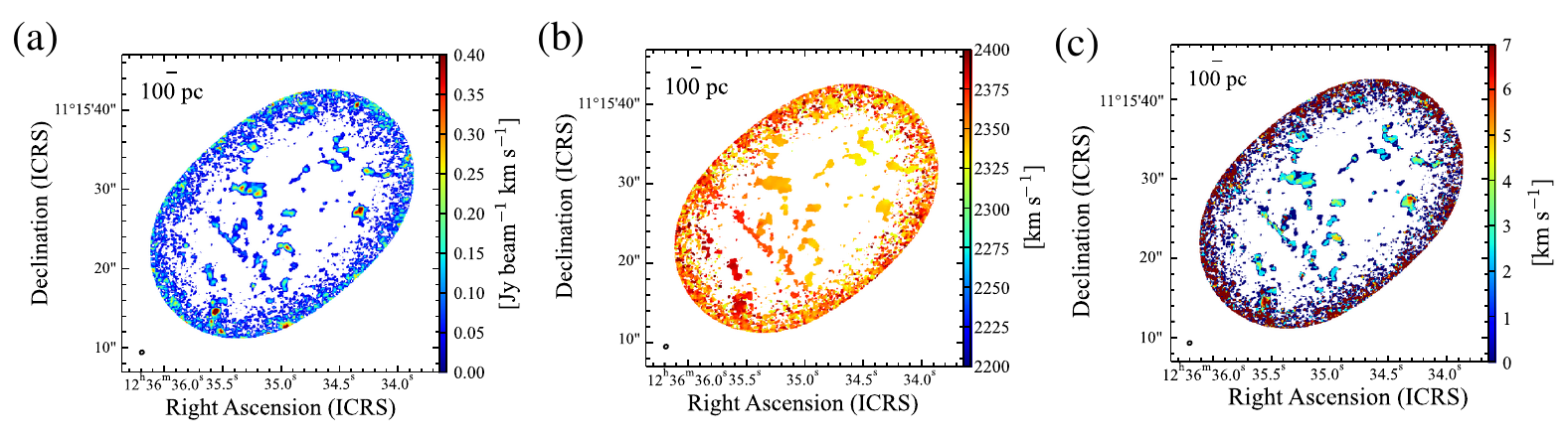}
	\end{center}
	\caption{CO($J$ = 3--2) moment maps. (a): Integrated intensity map. (b): Velocity field map. right: Velocity dispersion map. The beam size of \timeform{0.56"}$\times$\timeform{0.47"} and the scale bar are shown in the bottom-left corner and top-left corner, respectively.}
	\label{co32_moment_map}
\end{figure*}

The moment maps of CO($J$ = 3--2) are shown in figure \ref{co32_moment_map}. 
The integrated intensity map was clipped at 3\,$\sigma$, whereas the velocity field and the velocity dispersion map were clipped at 5\,$\sigma$ (1\,$\sigma$ = 1.5 mJy beam$^{-1}$).
Note that the primary beam correction caused noise enhancement at the edge of the field of view.
Multiple intensity peaks were observed in the field of view, suggesting the existence of warm and dense molecular gas.
The CO($J$ = 3--2) peaks trace the spiral arms of both galaxies. 
The position of the CO($J$ = 3--2) peaks in the overlapping region is not in alignment with the extension of the spiral arms of NGC\,4567, suggesting that the molecular gas may have been created by compression there. 
The location of such peaks may be the sites of future star formation.
The CO($J$ = 3--2) intensity shows a filamentary structure as the CO($J$ = 1--0) intensity does, although the CO($J$ = 3--2) intensity does not peak at the filament but around it.
The CO($J$ = 3--2) integrated intensity is 7.4\,$\pm$\,1.0 Jy km s$^{-1}$.
Within the CO($J$ = 3--2) field of view, the CO($J$ = 1--0) integrated intensity is 30$\pm$1 Jy km s$^{-1}$.
Therefore, the global CO($J$ = 3--2)/CO($J$ = 1--0) intensity ratio is 0.25\,$\pm$\,0.03.
		
\begin{figure}
	\begin{center}
		\includegraphics[width=\linewidth]{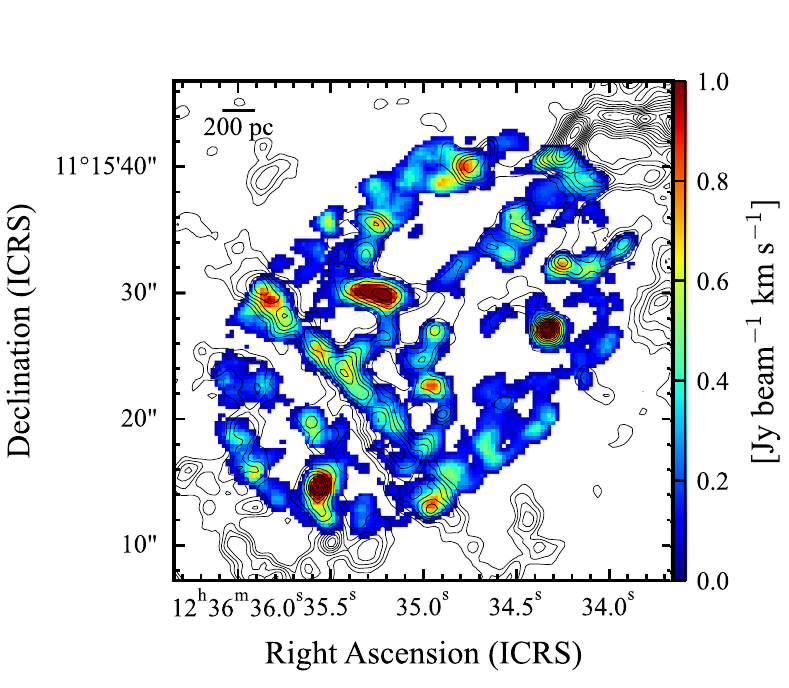}
	\end{center}
	\caption{CO($J$ = 3--2) integrated intensity map. Contours show the CO($J$ = 1--0) integrated intensity (The contour levels are 3\,$\sigma\,\times$\,1, 2, 3, ...).}
	\label{co32_mom0_map}
\end{figure}

Figure \ref{co32_mom0_map} shows the CO($J$ = 3--2) image overlaid on the CO($J$ = 1--0) contours, after convolving and re-gridding the CO($J$ = 3--2) data to match the lower resolution CO($J$ = 1--0) image.
In general, the distribution of the CO($J$ = 3--2) and CO($J$ = 1--0) exhibit consistency, including the locations of the strong peaks.
However, the intensity of CO($J$ = 3--2) along the filament is lower compared to other CO($J$ = 3--2) peaks, suggesting a deficiency of warm and dense molecular gas in the filament as opposed to the surrounding medium.
	
\subsection{Position-Velocity Diagram}\label{sec:pvd}
\begin{figure*}
	\begin{center}
		\includegraphics[width=\linewidth]{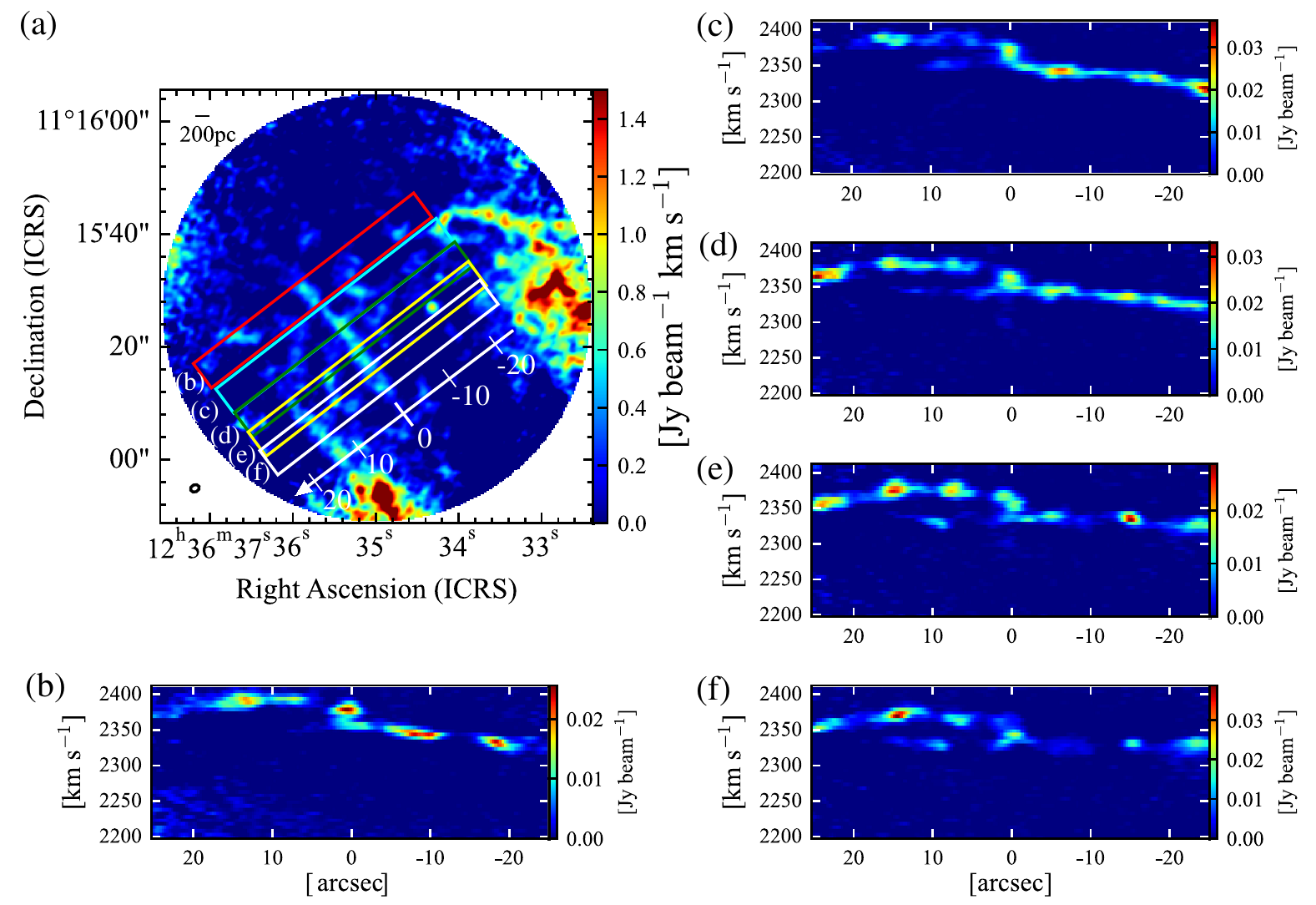}
	\end{center}
	\caption{(a): The CO(1-0) integrated intensity image with boxes that indicate the region where the position-velocity diagrams are made. The boxes of (b), (c), (d), (e), and (f) are centred on ($\alpha$, $\delta$)$_\mathrm{ICRS}$ = (\timeform{12h36m36.98s}, \timeform{11d15m12.72s}), (\timeform{12h36m36.71s}, \timeform{11d15m8.21s}), (\timeform{12h36m36.49s}, \timeform{11d15m4.05s}), (\timeform{12h36m36.32s}, \timeform{11d15m059s}), and (\timeform{12h36m36.18s}, \timeform{11d14m57.29s}), respectively. The long side of each box is perpendicular to the filament. (b)--(f): Position-velocity diagrams along the boxes shown in (a). The horizontal lines are offsets from the position of the filament.}
	\label{pvd}
\end{figure*}

In order to investigate the distribution and kinematics of the molecular gas in the overlapping region, we created position-velocity (P--V) diagrams that are perpendicular to the filament using CO($J$ = 1--0) data. 
Figure \ref{pvd} shows the specific areas where the P--V diagrams are made, along with the P--V diagram of each region.
Each region has a length of \timeform{50"} and a width of \timeform{5.4"}. 
In figure \ref{pvd}(b)--(f), the position of 0 arcsec corresponds to that of the filament. 
	
Each P--V diagram reveals that the 2300--2350 km s$^{-1}$ components from -20 to 0 arcsec arise from the emission from NGC\,4567, while the 2350--2400 km s$^{-1}$ emission from 0 to 20 arcsec comes from the disc of NGC\,4568.
The existence of molecular gas with a large-velocity width between those components is confirmed.
It is assumed to be evidence of a collision between the molecular gas of NGC\,4567 and NGC\,4568 \citep{Kaneko18}, indicating that the velocity difference is only about 50 km s$^{-1}$ for the line-of-sight direction.
Each P--V diagram shows weak emission at (velocity, offset) = ($\sim$2350 km s$^{-1}$, $\sim$10 arcsec), which is newly discovered in the high angular resolution data presented here.
The velocity of these components are connected with the disc of NGC\,4567, as seen in figure \ref{co10_moment_map}(b).
Since these components are not connected with the disc of NGC\,4568 in velocity, the extended molecular disc of NGC\,4567 at this position is not experiencing direct collision with the northern molecular disk of NGC\,4568.
This fact suggests that the colliding interface is limited to a small area around the filament, at least for molecular gas.
The small area of the collision front is consistent with the slight difference in the disc inclination of the two galaxies.
If the inclination of the two disks were the same, the collision interface would likely be broader and more violent.
The 0-arcsec component in figure \ref{pvd_filament}(f) shows that most of the molecular gas in that position belongs to the disc of NGC\,4567.
Therefore, the peculiar velocity component found in the southwest part of the filament ($\sim$2325 km s$^{-1}$; see section \ref{CO10}) can be explained as an extension of the spiral molecular arm emanating from the disc of NGC 4567, as suggested in section \ref{CO10}.

\section{Discussion}
\subsection{Molecular cloud identification}
\begin{figure*}
	\begin{center}
		\includegraphics[width=\linewidth]{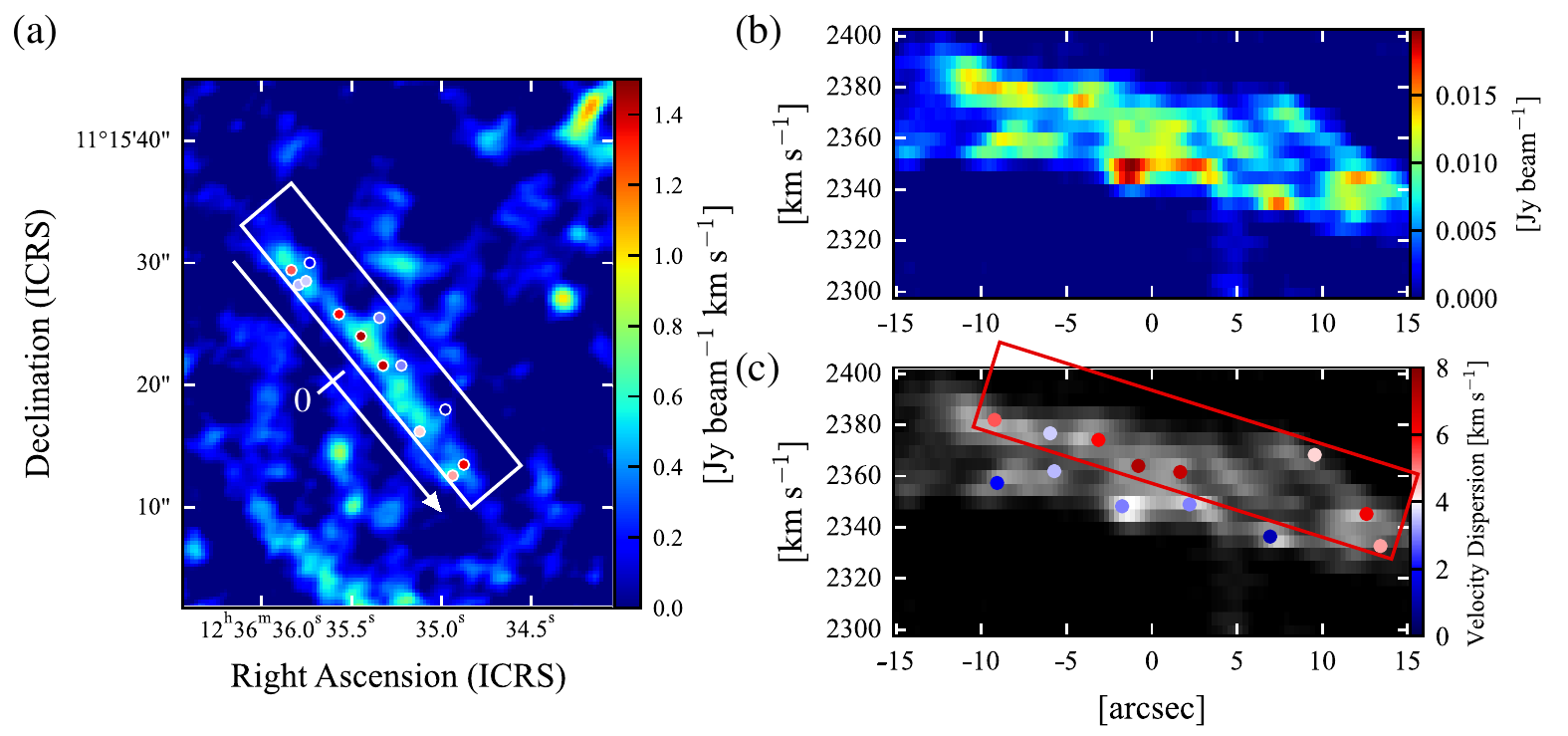}
	\end{center}
	\caption{(a) A close-up view of CO($J$ = 1--0) integrated intensity map around the filament. The white box is the area where the P--V diagram is drawn in (b) and (c). The positions of the molecular clouds identified in the region are shown and colour-coded by velocity dispersion. The colour of the velocity dispersion is the same as (c). (b) P--V diagram along the filament. (c) The same figure as (b), but the identified molecular clouds found in the same region are added. The colour corresponds to the velocity dispersion of each molecular cloud. The red box indicates the molecular gas that is considered to compose the filament.}
	\label{pvd_filament}
\end{figure*}

\begin{figure}
	\begin{center}
		\includegraphics[width=\linewidth]{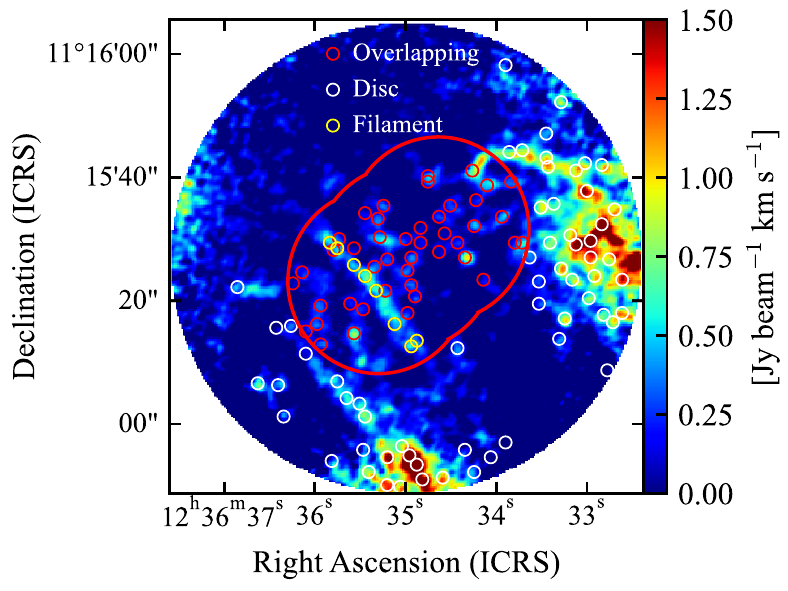}
	\end{center}
	\caption{Molecular clouds identified by CLUMPFIND (circle) overlaid on the CO($J$ = 1--0) integrated intensity map. Molecular clouds in the overlapping region (inside the field of view of CO($J$ = 3--2)) are shown as red circles. Eight yellow circles indicate molecular clouds associated with the filament. White circles correspond molecular clouds in the disc region (outside the field of view of CO($J$ = 3--2)).}
	\label{clumpfind_plot}
\end{figure}

To examine the physical states of molecular clouds in the overlapping region of NGC\,4567/4568, we identify molecular clouds using 3D cube data of CO($J$ = 1--0).
The CLUMPFIND algorithm \citep{Williams94} is employed for this purpose, searching for peaks from a given minimum level increasing with a given interval in a 3D cube.
We set the minimum level of contour lines at 3\,$\sigma$ and the interval of contour lines at 2\,$\sigma$. 
Since the primary beam correction has a substantial impact on the rms noise within the field of view of CO($J$ = 1--0), we separated the area into the overlapping region (inside the field of view of CO($J$ = 3--2)) and the disc region (outside the field of view of CO($J$ = 3--2)). 
The rms noise of 1\,$\sigma$ = 1.5 mJy beam$^{-1}$ is used for the overlapping region, and 1\,$\sigma$ = 4.0 mJy beam$^{-1}$ is used for the disc region.
A molecular cloud is defined as having a size larger than the beam size ($\sim$100 pc) and a velocity width larger than the velocity resolution (5 km s$^{-1}$).
To obtain precise physical quantities of molecular clouds, we exclude those truncated at the edge of the field of view.
In order to identify the number of molecular clouds associated with the filament, we plot the molecular clouds on the P--V diagram (figure \ref{pvd_filament}) and applied two criteria.
Firstly, the position of the identified cloud should be within the filament in the CO($J$ = 1--0) integrated intensity map.
Secondly, the systematic velocity of the cloud should be within the filament in the P--V diagram. 
The filament is defined as the area with CO($J$ = 1--0) integrated intensity of $>$\,0.5 Jy km s$^{-1}$ in the white box of figure \ref{pvd_filament}(a) and CO($J$ = 1--0) emission above 4.5 mJy beam$^{-1}$ (i.e., 3\,$\sigma$) in the P--V diagram of figure \ref{pvd_filament}(b).
As a result, 62 molecular clouds are identified in the disc region, 44 molecular clouds in the overlapping region (excluding those in the filament), and eight molecular clouds associated with the filament.
Figure \ref{pvd_filament}(c) shows a similar P--V diagram as in figure \ref{pvd_filament}(b), but the identified molecular clouds plotted in figure \ref{pvd_filament}(a) are overlaid. 
The eight molecular clouds surrounded by the red box in figure \ref{pvd_filament}(c) belong to the filament.
The five molecular clouds plotted outside the red box in figure \ref{pvd_filament}(c) are not identified as a molecular cloud in the filament, based on our adopted criteria described above.
The new high resolution and sensitivity CO($J$ = 1--0) map revealed four additional molecular clouds in the filament compared to the previous study by \citet{Kaneko18} who identified a total of four molecular clouds in the same region.
This difference can be explained by the higher angular resolution and sensitivity of our CO($J$ = 1--0) data.
Hereafter, we exclude `the molecular clouds in the filament' from `the molecular clouds in the overlapping region' in comparing the physical quantities.  
Figure \ref{clumpfind_plot} illustrates the position of the identified molecular clouds on the CO($J$ = 1--0) integrated intensity image.

\subsection{Physical properties of molecular clouds}\label{sec:phys_clouds}
\begin{figure*}
	\begin{center}
		\includegraphics[width=155mm]{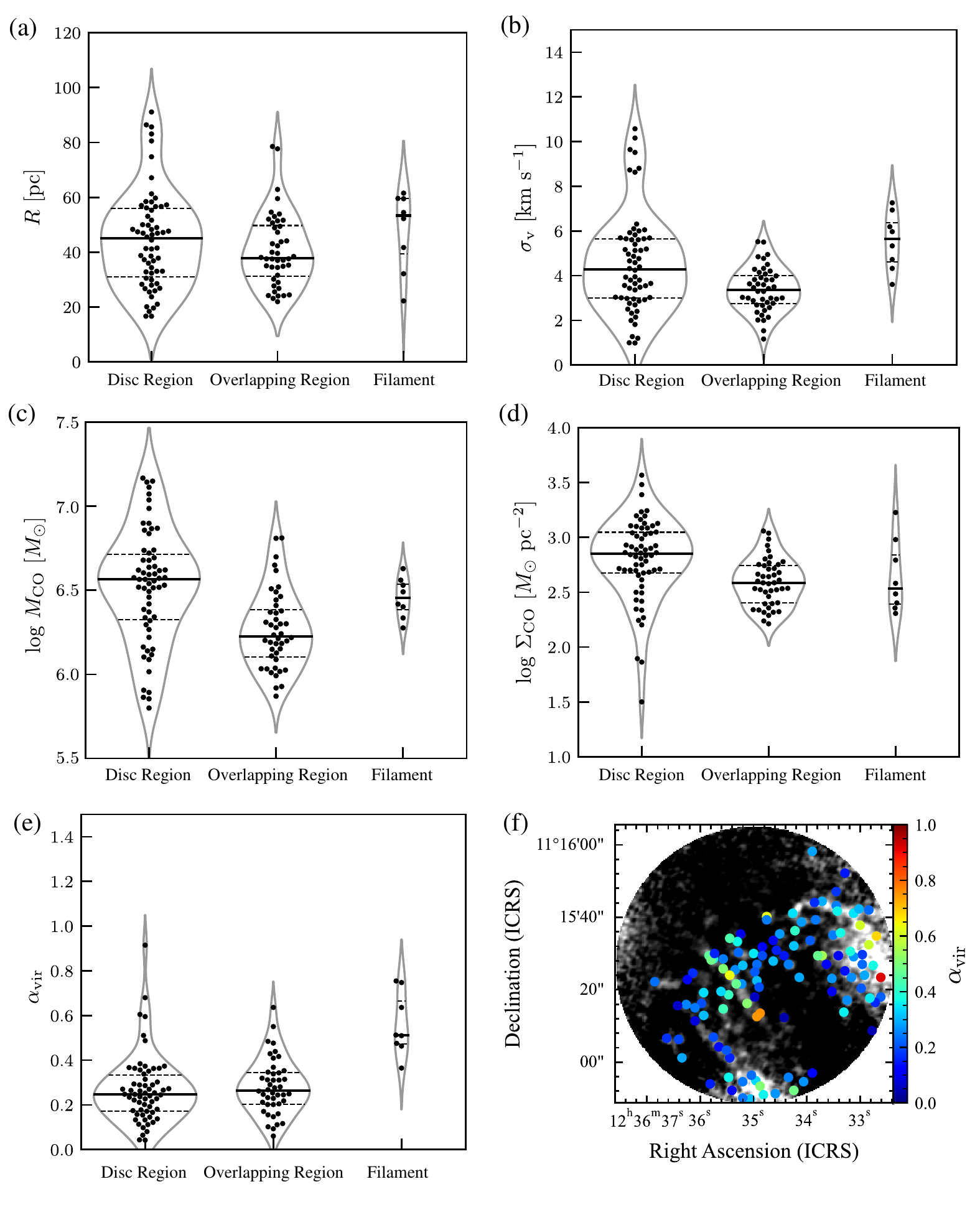}
	\end{center}
	\caption{(a): Violin plots of the deconvolved radius of the identified molecular clouds in each region. The bold horizontal line indicate the average value of each region. The upper and lower dashed horizontal lines show 75\% and 25\% percentiles, respectively. (b): Same as (a) but the violin plot of the velocity dispersion (c): Same as (a) but the violin plot of the luminosity mass in a logarithmic scale (d): Same as (a) but the violin plot of the mass surface density in a logarithmic scale. (e): Same as (a) but the violin plot of the virial parameter. (f): Positions of the identified molecular clouds. The colour indicates the virial parameter. The background grey-scale image is the CO($J$ = 1--0) integrated intensity.}
	\label{clumpfind_violin}
\end{figure*}

We derived the size (radius), velocity dispersion, luminosity mass, mass surface density, and virial parameter $\alpha_{\rm vir}$ for each identified molecular cloud. 
Radius and velocity dispersion were calculated after deconvolution during CLUMPFIND. 
The luminosity mass estimated from the CO integrated intensity was calculated using the following equation:
\begin{eqnarray}
	\left( \frac{M_{\rm CO}}{M_{\odot}} \right) = 2 && \left(\frac{m_{\rm H}}{M_{\odot}} \right) \left(\frac{A}{{\rm cm}^{2}} \right) \nonumber \\
	&& \left(\frac{X_{\rm CO}}{{\rm cm}^{-2} ({\rm K km s^{-1}})^{-1}} \right) \left(\frac{S_{\rm CO}}{{\rm K km s}^{-1}} \right),
\end{eqnarray}
where $m_{\rm H}$ is the hydrogen atom mass, $A$ is a projected area of a cloud, $X_{\rm CO}$ is a CO-to-H$_{2}$ conversion factor, and $S_{\rm CO}$ is the CO($J$ = 1--0) integrated intensity of each cloud. 
We adopt $X_{\rm CO}$ = 2.0\,$\times$\,10$^{20}$ [cm$^{-2}$ (K km s$^{-1})^{-1}$] \citep{Bolatto13} as in section \ref{CO10}.
Mass surface density was also calculated by dividing the luminosity mass by the area, assuming that the molecular cloud is a sphere.
We calculated the virial parameter $\alpha_{\rm vir}$ for the clouds using the following equation \citep{Bertoldi92}:
\begin{equation}
	\alpha_{\rm vir} = \frac{5\sigma^{2}R}{GM_{\rm CO}},
\end{equation}
where $R$ is a radius in pc, $\sigma$ is a velocity dispersion in km s$^{-1}$, $G$ is the gravitational constant, and $M_{\rm CO}$ is a luminosity mass in \MO.
Figure \ref{clumpfind_violin} shows violin plots of radius, velocity dispersion, luminosity mass, and mass surface density for each region. 
The typical uncertainties in the luminosity mass and the virial parameter are estimated to be $\sim$10\%, mainly due to the flux uncertainty.
Table \ref{Table2} shows the average values of the physical quantities of the molecular clouds in each region.

\begin{table*}
	\tbl{Properties of molecular clouds}{
		\begin{tabular}{lccccc}
			\hline
			Region & Radius & Velocity dispersion & Luminosity mass & Mass surface density & Virial parameter \\
			(Numbers of clouds) & [pc] & [km s$^{-1}$] & [$10^{6}$\,\MO] & [$10^{2}$\,\MO pc$^{-2}$] &  \\ \hline
			Disc (62)        & 45\,$\pm$\,18 & 4.6\,$\pm$\,2.3 & 4.5\,$\pm$\,3.5 & 8.5\,$\pm$\,6.5 & 0.26\,$\pm$\,0.16 \\
			Overlapping (44) & 41\,$\pm$\,13 & 3.4\,$\pm$\,1.0 & 2.2\,$\pm$\,1.4 & 4.5\,$\pm$\,2.0 & 0.28\,$\pm$\,0.12 \\
			Filament (8)     & 48\,$\pm$\,15 & 5.5\,$\pm$\,1.3 & 3.0\,$\pm$\,0.8 & 6.0\,$\pm$\,5.0 & 0.56\,$\pm$\,0.14 \\
			\hline
	\end{tabular}}
	\label{Table2}
\end{table*}
	
First, we examine the difference between the disc region and the overlapping region.
From figure \ref{clumpfind_violin}, while the radius and virial parameter do not show a clear difference between the disc region and overlapping region, there are differences in the velocity dispersion, luminosity mass, and mass surface density between the two regions.
The Kolmogorov--Smirnov (K--S) test is used to determine whether the population of the distribution of each physical quantity of the molecular clouds in the overlapping region and disc region is the same (the significance level is 1\%). 
The {\it p}-values of the K--S test for the radius and virial parameters above 1\%, specifically 0.23 and 0.67, respectively.
On the contrary, the {\it p}-values for the velocity dispersion, luminosity mass and mass surface density are significantly below 1\%, precisely 1.9\,$\times$\,10$^{-3}$, 3.1\,$\times$\,10$^{-6}$, and 2.8\,$\times$\,10$^{-5}$, respectively.
Compared to the overlapping region, the molecular gas in the disc region is distributed along the spiral arms and has a larger luminosity mass.
This is consistent with the scenario that molecular clouds in the spiral arm collide with each other, resulting  in larger mass, size, and velocity dispersion \citep{Egusa11,Hirota11,Sun20}.
	
Next, we investigate the molecular clouds in the filament. 
The clouds identified in the filament have larger velocity dispersion and hence, a factor of two higher $\alpha_{\rm vir}$ (0.56\,$\pm$\,0.14), compared to the surrounding clouds in the overlapping region (Figure \ref{clumpfind_violin}(f)).
If a molecular cloud has $\alpha_{\rm vir} \ltsim 1$, the cloud is considered to be gravitationally bound \citep{Kauffmann13}.
The derived $\alpha_{\rm vir}$ in the filament is a factor of three higher than that in \citet{Kaneko18} (0.18\,$\pm$\,0.10).
Considering the dispersion, the molecular clouds in the filament in this work have slightly larger radius and velocity dispersion than their data (their radius and velocity dispersion are 30.8\,$\pm$\,5.8 pc and 3.3\,$\pm$\,2.0 km s$^{-1}$, respectively).
We discover that the number of the identified clouds in the overlapping region is about a factor of two larger than \citet{Kaneko18}.
We presume that the difference in cloud identification is due to sensitivity since higher sensitivity may measure the size and velocity dispersion of molecular clouds that is diffuse in their outer edge more precisely. 
If it is the case, luminosity mass is roughly unchanged but derived size and velocity dispersion become larger.
It leads to larger virial parameters in our result in addition to the number of the identified clouds.
We conclude that although the molecular clouds in the filament have a larger $\alpha_{\rm vir}$ during the interaction, they are still gravitationally bound.
	
\begin{figure}
	\begin{center}
		\includegraphics[width=80mm]{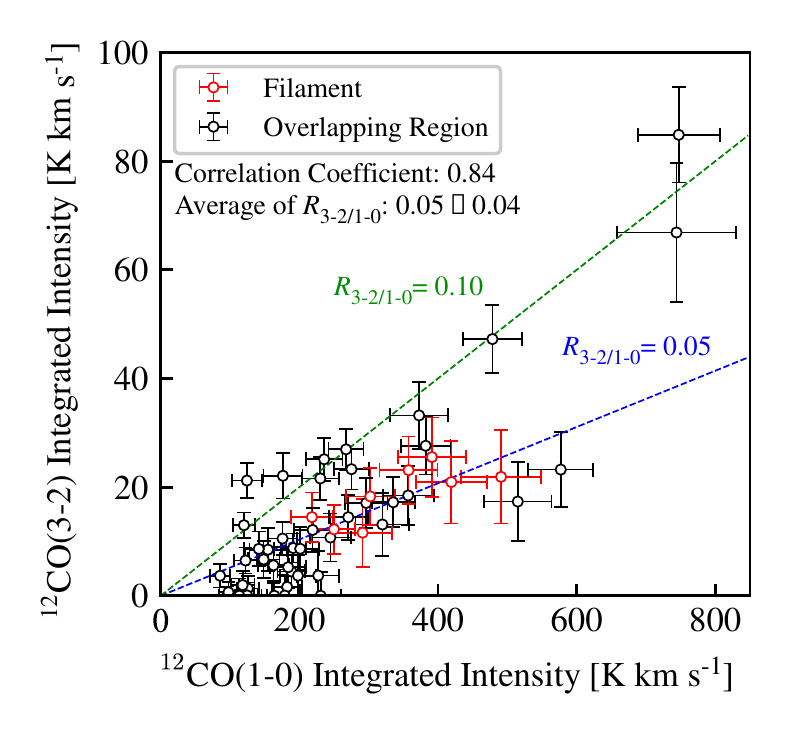}
	\end{center}
	\caption{Relation between the integrated intensity of CO($J$ = 1--0) and CO($J$ = 3--2) in each molecular cloud. Red: molecular clouds associating with the filament. Black: molecular clouds in the overlapping region. Blue and green dotted lines indicate that $R_{3-2/1-0}$ is 0.05 and 0.10, respectively.}
	\label{clumpfind_lineratio}
\end{figure}
	
Finally, we calculate $R_{3-2/1-0}$ on the molecular cloud scale. 
Regions where star formation occurs are expected to have high $R_{3-2/1-0}$.
For example, \citet{Lamperti20} found that LIRGs, star-forming regions, and AGN have an average $R_{3-2/1-0}$ of 0.5.
\citet{Saito15} showed that $R_{3-2/1-0}$ in the overlapping region of VV\,114, which is interacting galaxies displaying starbursts in the mid stage, has a $R_{3-2/1-0}$ value of 0.2--0.8.
On the other hand, the molecular clouds in the disc of the normal spiral galaxy M\,33 have an average $R_{3-2/1-0}$ of 0.26, measured only using CO($J$ = 1--0) peak pixels \citep{Onodera12}.
In this study, we calculate the integrated intensity ratios using the pixel values in the molecular cloud identified using the CO($J$ = 1--0) cube data.
Figure \ref{clumpfind_lineratio} shows the relationship between the integrated intensity of CO($J$ = 1--0) and CO($J$ = 3--2) in each molecular cloud.
The clouds in the disc region are not plotted in figure \ref{clumpfind_lineratio}.
The average $R_{3-2/1-0}$ for each molecular cloud in the overlapping region is 0.05\,$\pm$\,0.04.
All molecular clouds in the filaments have an $R_{3-2/1-0}$ of less than 0.1, and the maximum value of $R_{3-2/1-0}$ for all molecular clouds, including the overlapping region, is 0.17\,$\pm$\,0.04.
The correlation coefficient between CO($J$ = 1--0) and CO($J$ = 3--2) integrated intensity of each molecular cloud is 0.84, indicating that all molecular clouds contain a fairly constant fraction of warmer and dense gas.
Compared to the previous studies, the molecular clouds identified in this study have significantly lower $R_{3-2/1-0}$.
These line ratios are smaller than those for GMCs in the disc of the Milky Way (0.4-0.5: \cite{Sanders93,Oka07}). 
Furthermore, $R_{3-2/1-0}$ in the overlapping region is comparable to the ratio of 0.10 found in the inter-arm region of NGC\,613 \citep{Muraoka16}.
Therefore, the environment of the overlapping region may be significantly colder and more diffuse than that of the disc of spiral galaxies.
Low $R_{3-2/1-0}$ also suggests that molecular gas compression is not efficiently induced at this merging stage.

\begin{figure}
	\begin{center}
		\includegraphics[width=\linewidth]{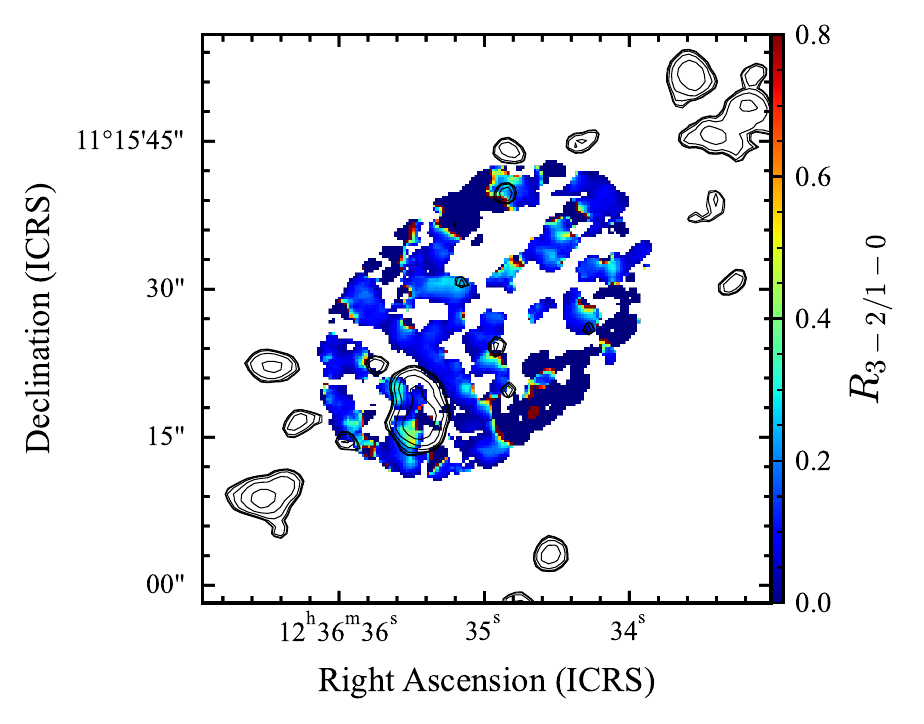}
	\end{center}
	\caption{$R_{3-2/1-0}$ image with H$\alpha$ contours. The contour levels are 2.5\,$\times$\,10$^{n}$ erg s$^{-1}$, where $n$ = 1, 2, 3, ....}
	\label{fig:R32-Ha}
\end{figure}

Given the low $R_{3-2/1-0}$ but larger $\alpha_{\rm vir}$ than the surrounding region, it is improbable for star formation to occur soon within the filament of the overlapping region of this galaxy pair.
To investigate this possibility, we check the distributions of $R_{3-2/1-0}$ and a star formation tracer, H$\alpha$ \citep{Xu00}.
Figure \ref{fig:R32-Ha} shows the $R_{3-2/1-0}$ image overlaid with H$\alpha$.
Note that since the $R_{3-2/1-0}$ image is derived from CO($J$ = 3--2) and CO($J$ = 1--0) integrated over the line of sight, there may be emission that is not associated with the molecular clouds.
Therefore, the values presented in figure \ref{fig:R32-Ha} cannot be directly compared with those in figure \ref{clumpfind_lineratio}. 
The $R_{3-2/1-0}$ in the bulk of the filament is lower than 0.2 except for some parts of the outer edge region.
Many of the star-forming regions traced by H$\alpha$ emission are associated with high $R_{3-2/1-0}$ ($R_{3-2/1-0}\,\gtrsim\,0.4$).
However, we also identify regions with high $R_{3-2/1-0}$ that are not associated with any H$\alpha$ emissions, which may become future sites for star formation.

\subsection{Formation process of the filamentary molecular structure}
In this section, we investigate the formation process of the filament.
As discussed in section \ref{sec:phys_clouds}, our data illustrate that molecular clouds in the filament have low $R_{3-2/1-0}$, which is similar to the values in the overlapping region on average.
This means that they are still diffuse although they have a slightly larger mass than that in the overlapping region.
However, \citet{Saitoh09} demonstrate that the molecular filament becomes dense enough to form stars within a few tens of Myr after its formation.
These facts suggest that the molecular filament in NGC\,4567/4568 is a good target to investigate the origin of the filament rather than dense gas formation during the interaction, and moderately dense gas tracers such as CO($J$ = 3--2) may not be the ideal choice to probe the formation process.
Here we explore the possibility that the filament is formed through the compression of diffuse gas along the collision front. 
For this purpose, we compare the galaxy-scale distributions between H\emissiontype{I} and CO, and investigate their properties, including how the intercluster medium (ICM) may have affected the formation of the filament in this galaxy pair located in the Virgo cluster.
	
\begin{figure*}
	\begin{center}
		\includegraphics[width=\linewidth]{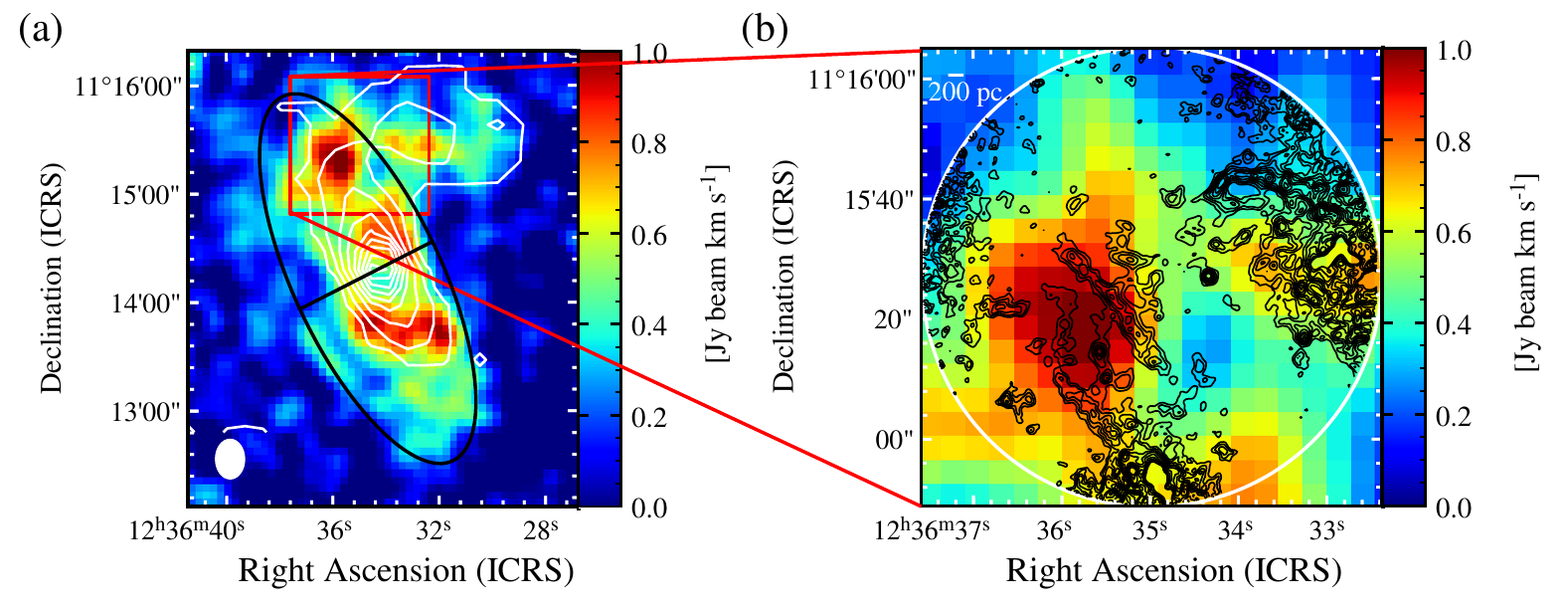}
	\end{center}
	\caption{(a) Colour: H$\emissiontype{I}$ integrated intensity map observed by VLA \citep{Iono05}. 
		Contour: CO($J$ = 1--0) integrated intensity map obtained with the Nobeyama 45-m telescope \citep{Kaneko13}. 
		The contour levels are 7.20\,$\times$\,1, 2, 3, ... K km s$^{-1}$. 
		The beam size of H$\emissiontype{I}$ is shown as a white ellipse in the bottom-left corner.
		The upper side of a black eclipse is the northern disc region of NGC\,4568 and the bottom is the southern disc region of NGC\,4568 (the definition is written in the text). 
		(b) Colour: Close-up view of H$\emissiontype{I}$ gas. Contour: CO($J$ = 1--0) integrated intensity map obtained with the ALMA Cycle 3. 
		The contour levels are 0.14\,$\times$\,1, 2, 3, ... Jy beam$^{-1}$ km s$^{-1}$. 
		The white circle indicates the field of view of CO($J$ = 1--0) of the ALMA Cycle 3 observations.}
	\label{hi_co}
\end{figure*}

Figure \ref{hi_co}(a) shows an H$\emissiontype{I}$ image \citep{Iono05} overlaid with CO($J$ = 1--0) contours obtained by the Nobeyama 45-m telescope \citep{Kaneko13}.
In NGC\,4567, the extent of H\emissiontype{I} gas is comparable to that of CO.
In contrast, H$\emissiontype{I}$ gas in the southern disc of NGC\,4568 is distributed outside CO gas, which is commonly seen in isolated galaxies.
On the other hand, the H$\emissiontype{I}$ extent in the northern disc of NGC\,4568 is the same as the CO($J$ = 1--0) extent.
In addition, the peak of H$\emissiontype{I}$ gas in the northern disc of NGC\,4568 is located about 500\,pc east of the centre of the filament.
	
First, we investigate whether H\emissiontype{I} gas is affected by the potential of the Virgo cluster.
\citet{Chung09} estimated that NGC\,4567 and NGC\,4568 have H\emissiontype{I} deficiency of 0.13\,$\pm$\,0.12 and 0.38\,$\pm$\,0.12, respectively.
Considering the scatter and uncertainty in estimating the H\emissiontype{I} gas mass of isolated galaxies \citep{Giovanelli83}, these values indicate weak or no H\emissiontype{I} removal by the ICM.
Furthermore, \citet{Yoon17} categorised galaxies in the Virgo cluster into five groups based on their morphology, escape velocity, and H\emissiontype{I} deficiency.
Both NGC\,4567 and NGC\,4568 are categorised as Class 0, which means no clear sign of gas stripping by the ICM.
In order to check the H$\emissiontype{I}$ gas stripping, we divided NGC\,4568 into two regions: the northern disc and the southern disc.
We perform an eclipse fit to the region containing NGC\,4568 for the H\emissiontype{I} image to reproduce the 3$\sigma$ level. 
The eclipse is then split along the minor axis of NGC\,4568. 
The northern disc region is defined as the side containing the overlap region, and the south disc region is defined as the remaining side.
These regions are depicted in black in figure \ref{hi_co}(a). 
Note that the northern disc region includes the overlapping region.
Assuming the optically thin emission, the H$\emissiontype{I}$ gas mass in the northern disc region of NGC\,4568 (($5.6\,\pm\,0.3)\,\times\,$10$^{8}$\,\MO) is comparable to that in the southern side (($4.9\,\pm\,0.2)\,\times$\,10$^{8}$\,\MO).
This implies that H\emissiontype{I} gas is not stripped in the overlapping region by the ICM, although it shows an asymmetric H\emissiontype{I} distribution.
These facts imply that the cluster environment may affect NGC\,4567 and NGC\,4568 but not strongly compared to other severely affected cluster galaxies, which show high H\emissiontype{I} deficiency, H\emissiontype{I} truncation, or warped discs.
The environmental effect may not be the main driver for making the filament.
	
We consider the possibility that the filament is made by the interaction.
Figure \ref{hi_co}(b) illustrates the close-up view of H$\emissiontype{I}$ and CO($J$ = 1--0) obtained with the ALMA Cycle 3 program in the overlapping region.
Molecular filament is located around the H$\emissiontype{I}$ peak.
Recent simulation suggests that fast collisions of H$\emissiontype{I}$ gas, such as under a galaxy interaction, can make massive molecular clumps ($>$\,10$^{4}$\,\MO) in a short time due to shock compression \citep{Maeda21}.
\citet{Kaneko17} showed that the overlapping region has a slightly higher molecular gas fraction than the region with a similar surface density, suggesting an effective transition from atomic gas to molecular gas.
Thus, the shock compression of H$\emissiontype{I}$ may form the filament at the collision front.
The constant $R_{3-2/1-0}$ ratio within the filament implies that this compression process does not occur randomly at the collision front, but occurs almost simultaneously.
High-resolution H$\emissiontype{I}$ observation is necessary to verify this possibility.

\subsection{Evolutionary and collisional stages of molecular clouds in NGC\,4567/4568}
\begin{figure*}
	\begin{center}
		\includegraphics[width=\linewidth]{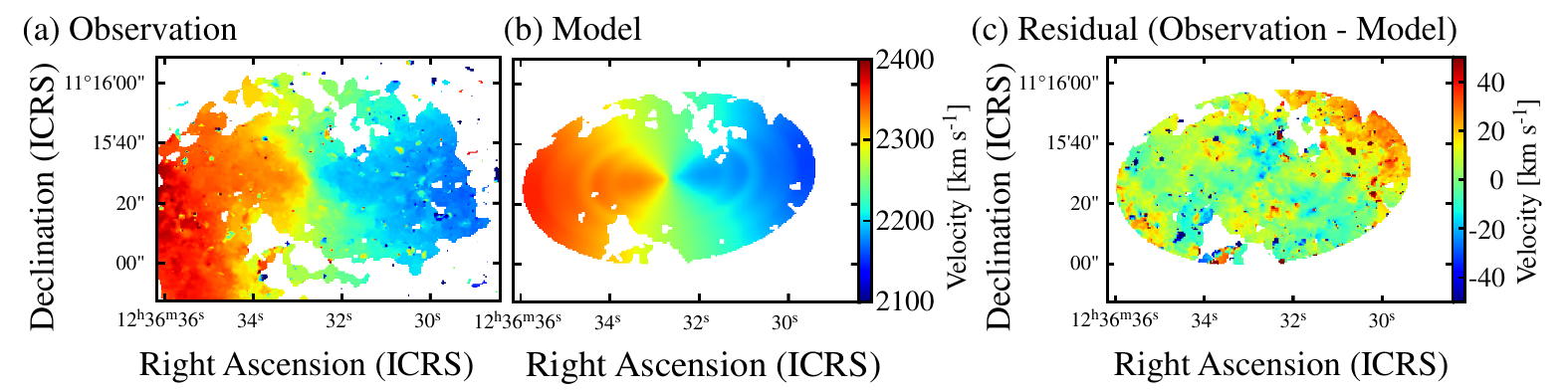}
	\end{center}
	\caption{Velocity fields for NGC\,4567 obtained from the observation and the kinematic model created by 3D-BAROLO. (a) velocity field from the observation, (b) kinematic model, (c) residual (observation - kinematic model)}
	\label{barolo4567}
\end{figure*}

\begin{figure*}
	\begin{center}
		\includegraphics[width=\linewidth]{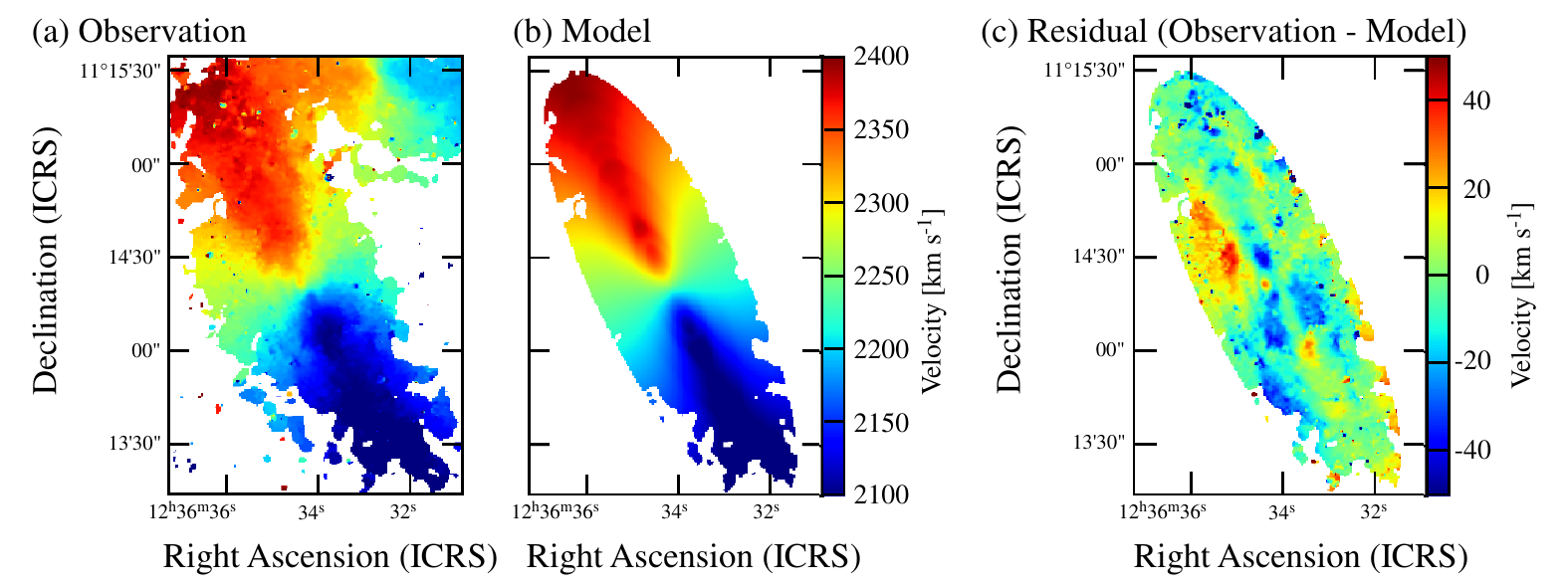}
	\end{center}
	\caption{Velocity fields for NGC\,4568 obtained from the observation and the kinematic model created by 3D-BAROLO. (a) velocity field from the observation, (b) kinematic model, (c) residual (observation - kinematic model)}
	\label{barolo4568}
\end{figure*}

Numerical simulations suggest that widespread star formation can occur during the early stage of the interaction, as compressive turbulence or shock make molecular gas dense \citep{Barnes04, Renaud15}.
Although the NGC\,4567/4568 pair does not show violent star formation, it is worth investigating whether such disturbances are occurring. 
For this purpose, we create kinematic models using 3D-BAROLO \citep{DiTeodoro15}, assuming a circular motion, and compare them with the observed data of each galaxy pair obtained in the ALMA Cycle 1 program (figures \ref{barolo4567} and \ref{barolo4568}).
We find slight distortion around (R.A., Decl.) = (\timeform{12h36m36s}, \timeform{+11d14'40''}), which may originate from the initial response of the disc to the interaction. 
Except for this distortion, there is no significant deviation from the circular motion, including the overlapping region, suggesting that the velocity structures of both galaxies are not significantly altered, and that the overall impact of the interaction is limited.
These results suggest that this galaxy pair is in a very early stage of interaction, as previously suggested by \citet{Kaneko13}.
Although the pair appears to collide approximately in parallel, the molecular discs collide in a small area, as shown in figure \ref{pvd_filament}.
A small effective collision area and short compression time by the interaction due to the oblique collision may be one of the reasons why the filament does not become dense.
	

\section{Summary}
We investigated the physical state of molecular gas in the overlapping region by observing CO($J$ = 1--0) and CO($J$ = 3--2) in an early-stage interacting galaxy pair NGC\,4567/4568 with ALMA.
The summary of this paper is as follows:

\begin{enumerate}
	\item Diffuse gas tracer CO($J$ = 1--0) and warmer and denser gas tracer CO($J$ = 3--2) are detected in the overlapping region of this galaxy pair.
	We confirm the existence of a filamentary structure of CO($J$ = 1--0) emission, which is likely tracing the collision front. 
	The peak of CO($J$ = 3--2) emission, on the other hand, is not at the filamentary structure.

	\item We identified 114 molecular clouds using CO($J$ = 1--0) data. 
	We compared the properties of the molecular clouds by dividing the field of view of CO($J$ = 3--2) into the overlapping region and the disc region.
	The $p$-values of the Kolmogorov-Smirnov test for the velocity dispersion, luminosity mass, and mass surface density are 1.9\,$\times$\,10$^{-3}$, 3.1\,$\times$\,10$^{-6}$, and 2.8\,$\times$\,10$^{-5}$. 
	Thus, the velocity dispersion, luminosity mass, mass surface density of molecular clouds in the disc region are larger than those in the overlapping region.

	\item Molecular clouds in the filament have larger velocity dispersion and a factor of two higher virial parameters. 
	However, all molecular clouds in the filament are gravitationally bound (virial parameter lower than 1).
	The higher virial parameter than the surrounding region indicates that it is due to turbulence induced by collisions of molecular gas and/or atomic gas.

	\item We calculated CO($J$ = 3--2)/CO($J$ = 1--0) intensity ratios for each identified molecular cloud. 
	The maximum CO($J$ = 3--2)/CO($J$ = 1--0) ratio is 0.17\,$\pm$\,0.04. 
	The mean value of CO($J$ = 3--2)/CO($J$ = 1--0) for each molecular cloud is 0.05\,$\pm$\,0.04, which is smaller than the values in star-forming regions of other interacting galaxies. 
	The molecular gas in the overlapping region does not become denser efficiently, suggesting that it is not at the stage star formation occurs.

	\item The molecular filament is located around the H\emissiontype{I} peak in the overlapping region, and the H\emissiontype{I} gas in this galaxy pair is most abundant in the overlapping region.
	The distributions of CO and H\emissiontype{I} suggest that the filament might have been formed by H\emissiontype{I} gas compression.

	\item We made the kinematic model for NGC\,4567 and NGC\,4568.
	Although we find a slight distortion in a part of the disc region of NGC\,4568, no significant distortion of the velocity structure for the entire pair is seen.
	The results indicate that this galaxy pair is at a very early stage of interaction and that a very quiet collision has occurred.
\end{enumerate}
Our results suggest molecular clouds in the overlapping region and the filament do not consist of dense gas. 
Since this galaxy pair is at the very early stage of interaction, molecular clouds in the overlapping region may become denser and start star formation along the progress of interaction. 
High-angular and velocity resolution H\emissiontype{I} observation would be a key to understanding the formation mechanism of the molecular filament.

\bigskip
\section*{Funding}
This work was supported by the NAOJ ALMA Scientific Research Grant Number 2020-15A.
H.K. was supported by the ALMA Japan Research Grant of NAOJ ALMA Project, NAOJ-ALMA-275.
	
\section*{Acknowledgements}
We would like to thank the anonymous referee for many constructive comments that helped to improve the manuscript.
This paper makes use of the following ALMA data: ADS/JAO.ALMA\#2015.1.01161.S and ADS/JAO.ALMA\#2012.1.00759.S. ALMA is a partnership of ESO (representing its member states), NSF (USA) and NINS (Japan), together with NRC (Canada) and NSC and ASIAA (Taiwan) and KASI (Republic of Korea), in cooperation with the Republic of Chile. 
The Joint ALMA Observatory is operated by ESO, AUI/NRAO and NAOJ. 
	
This research has made use of the NASA/IPAC Extragalactic Database (NED) which is operated by the Jet Propulsion Laboratory, California Institute of Technology, under contract with the National Aeronautics and Space Administration.
	
Data analysis was partly carried out on the common-use data analysis computer system at the Astronomy Data Center (ADC) of the National Astronomical Observatory of Japan.


\begin{thebibliography}{}
	\bibitem[Barnes(2004)]{Barnes04} Barnes, J.~E. \ 2004, \mnras, 350, 798
	\bibitem[Bertoldi \& McKee(1992)]{Bertoldi92} Bertoldi, F., \& McKee, C.~F. \ 1992, \apj, 395, 140
	\bibitem[Bolatto, Wolfire, \& Leroy(2013)]{Bolatto13} Bolatto, A.~D., Wolfire, M., \& Leroy, A.~K. \ 2013, \araa, 51, 207
	\bibitem[Bushouse(1986)]{Bushouse86} Bushouse, H.~A. \ 1986, \aj, 91, 255
	\bibitem[CASA Team \etal(2022)]{CASA22} CASA Team \etal \ 2022, \pasp, 134, 114501
	\bibitem[Chung \etal(2009)]{Chung09} Chung, A., van Gorkom, J.~H., Kenney, J.~D.~P., Crowl, H., \& Vollmer, B. \ 2009, \aj, 138, 1741
	\bibitem[Di Teodoro \& Fraternali(2015)]{DiTeodoro15} Di Teodoro, E.~M., \& Fraternali, F. \ 2015, \mnras, 451, 3021
	\bibitem[Egusa, Koda, \& Scoville(2011)]{Egusa11} Egusa, F., Koda, J., \& Scoville, N. \ 2011, \apj, 726, 85
	\bibitem[Gao \& Solomon(2004)]{Gao04} Gao, Y., \& Solomon, P.~M. \ 2004, \apj, 606, 271
	\bibitem[Garc{\'\i}a-Burillo \etal(2012)]{Garcia12} Garc{\'\i}a-Burillo, S. \etal \ 2012, \aap, 539, A8
	\bibitem[Giovanelli \& Haynes(1983)]{Giovanelli83} Giovanelli, R., \& Haynes, M.~P. \ 1983, \aj, 88, 881
	\bibitem[Graci{\'a}-Carpio \etal(2008)]{Gracia08} Graci{\'a}-Carpio, J., Garc{\'\i}a-Burillo, S., Planesas, P., Fuente, A., \& Usero, A. \ 2008, \aap, 479, 703
	\bibitem[Hirota \etal(2011)]{Hirota11} Hirota, A., Kuno, N., Sato, N., Nakanishi, H., Tosaki, T., \& Sorai, K. \ 2011, \apj, 737, 40
	\bibitem[Iono, Yun, \& Ho(2005)]{Iono05} Iono, D., Yun, M.~S., \& Ho, P.~T.~P. \ 2005, \apjs, 158, 1
	\bibitem[Kaneko \etal(2013)]{Kaneko13} Kaneko, H., Kuno, N., Iono, D., Tamura, Y., Tosaki, T., Nakanishi, K., \& Sawada T. \ 2013, \pasj, 65, 20
	\bibitem[Kaneko \etal(2017)]{Kaneko17} Kaneko, H., Kuno, N., Iono, D., Tamura, Y., Tosaki, T., Nakanishi, K., \& Sawada T. \ 2017, \pasj, 69, 66
	\bibitem[Kaneko \etal(2022)]{Kaneko22} Kaneko, H., Kuno, N., Iono, D., Tamura, Y., Tosaki, T., Nakanishi, K., \& Sawada T. \ 2022, \pasj, 74, 343
	\bibitem[Kaneko, Kuno, \& Saitoh(2018)]{Kaneko18} Kaneko H., Kuno N., Saitoh T.~R. \ 2018, \apjl, 860, L14
	\bibitem[Kauffmann, Pillai, \& Goldsmith(2013)]{Kauffmann13} Kauffmann, J., Pillai, T., \& Goldsmith, P.~F. \ 2013, \apj, 779, 185
	\bibitem[Kenney \& Young(1998)]{Kenney98} Kenney, J.~D., \& Young, J.~S. \ 1988, \apjs, 66, 261
	\bibitem[Kennicutt \etal(1987)]{Kennicutt87} Kennicutt, R.~C., Keel, W.~C., van der Hulst, J.~M., Hummel, E., \& Roettiger, K.~A. \ 1987, \aj, 93, 1011
	\bibitem[Knapen \& James(2009)]{Knapen09} Knapen J.~H., \& James P.~A. \ 2009, \apj, 698, 1437
	\bibitem[Lamperti \etal(2020)]{Lamperti20} Lamperti, I. \etal \ 2020, ApJ, 889, 103
	\bibitem[Maeda, Inoue, \& Fukui(2021)]{Maeda21} Maeda R., Inoue T., \& Fukui Y., \ 2021, \apj, 908, 2
	\bibitem[Mei \etal(2007)]{Mei07} Mei, S., \etal \ 2007, \apj, 655, 144
	\bibitem[Mengel \etal(2008)]{Mengel08} Mengel, S., Lehnert, M.~D., Thatte, N. A., Vacca, W. D., Whitmore, B., \& Chandar, R. \ 2008, \aap, 489, 1091
	\bibitem[Muraoka \etal(2016)]{Muraoka16} Muraoka, K., Takeda, M., Yanagitani, K., \etal \ 2016, PASJ, 68, 18
	\bibitem[Oka \etal(2007)]{Oka07} Oka, T., Nagai, M., Kamegai, K., Tanaka, K., \& Kuboi, N. \ 2007, \pasj, 59, 15
	\bibitem[Onodera \etal(2012)]{Onodera12} Onodera, S. \etal \ 2012, \pasj, 64, 133
	\bibitem[Renaud, Bournaud, \& Duc(2015)]{Renaud15} Renaud F., Bournaud F., \& Duc P.-A. \ 2015, \mnras, 446, 2038
	\bibitem[Saito \etal(2015)]{Saito15} Saito, T. \etal \ 2015, \apj, 803, 60
	\bibitem[Saitoh \etal(2009)]{Saitoh09} Saitoh T.~R., Daisaka H., Kokubo E., Makino J., Okamoto T., Tomisaka K., Wada K., \& Yoshida, N. \ 2009, \pasj, 61, 481
	\bibitem[Saitoh \etal(2011)]{Saitoh11} Saitoh T.~R. \etal \ 2011, in IAU Symp. 270, Computational Star Formation, ed. J. Alves \etal \ (Cambridge: Cambridge Univ. Press), 483
	\bibitem[Sanders \& Mirabel(1996)]{Sanders96} Sanders, D.~B., \& Mirabel, I.~F. \ 1996, \araa, 34, 749
	\bibitem[Sanders \etal(1999)]{Sanders93} Sanders, D. B., Scoville, N. Z., Tilanus, R. P. J., Wang, Z., \& Zhou, S. \ 1993, in AIP Conf. Proc., 278, Back to the Galaxy, ed. S. S. Holt \& F. Verter (New York: AIP), 311
	\bibitem[Sun \etal(2020)]{Sun20} Sun, J. \etal \ 2020, \apjl, 901, L8
	\bibitem[Teyssier, Chapon, \& Bournaud(2010)]{Teyssier10} Teyssier, R., Chapon, D., \& Bournaud, F. \ 2010, \apjl, 720, L149
	\bibitem[Vollmer \etal(2021)]{Vollmer21} Vollmer, B., Braine, J., Mazzilli-Ciraulo, B., \& Schneider, B. \ 2021, \aap, 647, 138
	\bibitem[Wiliams, de Geus, \& Blitz(1994)]{Williams94} Williams, J.~P., de Geus, E.~J., \& Blitz, L. \ 1994, \apj, 428, 693
	\bibitem[Wilson \etal(2000)]{Wilson00} Wilson, C.~D., Scoville, N., Madden, S.~C., \& Charmandaris, V. \ 2000, \apj, 542, 120
	\bibitem[Wilson \etal(2003)]{Wilson03} Wilson, C.~D., Scoville, N., Madden, S.~C., \& Charmandaris, V. \ 2003, \apj, 599, 1049
	\bibitem[Whitemore \etal(2014)]{Whitmore14} Whitmore, B.~C. \etal \ 2014, \apj, 795, 156
	\bibitem[Xu \etal(2000)]{Xu00} Xu, C., Gao, Y., Mazzarella, J., Lu, N., Sulentic, J.~W., Domingue, D.~L. \ 2000, \apj, 541, 644
	\bibitem[Yoon \etal(2017)]{Yoon17} Yoon, H., Chung, A., Smith, R., \& Jaff{\'e}, Y.~L., \ 2017, \apj, 838, 81
	\bibitem[Zhu \etal(2007)]{Zhu07} Zhu, M., Gao, Y., Seaquist, E.~R. \& Dunne, L. \ 2007, \aj, 134, 118
\end{thebibliography}
\end{document}